\documentclass[twocolumn,amsmath,amssymb,10pt,aps]{revtex4}
  
\usepackage[utf8]{inputenc}
\usepackage[T1]{fontenc}

\usepackage{amsmath}
\usepackage{amsfonts}
\usepackage{graphicx}
\usepackage{yfonts}
\usepackage{color}
\usepackage[normalem]{ulem}
\usepackage{amsthm}
\usepackage{bm}
\usepackage{bbm}
\usepackage{mathtools}
\usepackage{array}
\usepackage{placeins}
\usepackage{enumitem}

\newcommand{\der}{\mathrm{d}}

\def\<{\langle}
\def\>{\rangle}

\newcommand{\Tr}{\mathrm{Tr}}
\def\oper{{\mathchoice{\rm 1\mskip-4mu l}{\rm 1\mskip-4mu l}
{\rm 1\mskip-4.5mu l}{\rm 1\mskip-5mu l}}}
\DeclareMathAlphabet\mathbfcal{OMS}{cmsy}{b}{n}

\begin{document}

\title{Geometry of the Pauli maps and Pauli channels}

\author{Katarzyna Siudzi{\'n}ska}
\affiliation{Institute of Physics, Faculty of Physics, Astronomy and Informatics \\  Nicolaus Copernicus University, Grudzi\k{a}dzka 5/7, 87--100 Toru{\'n}, Poland}

\begin{abstract}
We analyze the geometrical properties of trace-preserving Pauli maps. Using the Choi-Jamio{\l}kowski isomorphism, we express the Hilbert-Schmidt line and volume elements in terms of the eigenvalues of the Pauli map. We analytically compute the volumes of positive, trace-preserving Pauli maps and Pauli channels. In particular, we find the relative volumes of the entanglement breaking Pauli channels, as well as the channels that can be generated by a time-local generator. Finally, we show what part of the Pauli channels are P and CP-divisible, which is related to the notion of Markovianity.
\end{abstract}

\flushbottom

\maketitle

\thispagestyle{empty}

\section{Introduction}

Recently, much attention has been given to the geometrical properties of quantum states. To determine the distance between two states, one introduces the metric, which determines the geometry of the underlying space. For pure states, there exists a unique unitarily invariant metric, which is induced by the Fubini-Study line element \cite{Zyczkowski}. However, for mixed states, there are several possible choices of the metric. In \cite{Vitale}, the authors derived a family of metrics for $N$-level quantum states from the relative entropy. Bures and \.{Z}yczkowski used the Hilbert-Schmidt metric \cite{Sommers2} and the Bures metric \cite{Sommers} to compute the volume of mixed quantum states in an arbitrary finite dimension. In the infinite-dimensional case, it was shown that three different metrics produce the same volume of the Gaussian states with a fixed global purity \cite{Link2015}.

Similarly, there are papers dedicated to analyze the geometry of quantum channels. Narang and Arvind \cite{Arvind} calculated the volume of the Pauli channels that can be simulated by a one-qubit mixed state environment. Analogical calculations were carried out for the generalized amplitude damping channels \cite{Jung}.
Due to the duality between quantum states and quantum maps, one can use the geometrical formalism developed for quantum states to study quantum maps. The one-to-one correspondence between states and channels is established by the Choi-Jamio{\l}kowski isomorphism \cite{Choi,JAMIOLKOWSKI1972275}. It has been applied by Lovas and Andai \cite{Lovas} to compute the volumes of the general and unital qubit channels using the Lebesque measure. Szarek et. al. \cite{Szarek2} derived bounds for the Hilbert-Schmidt volumes of positive, trace-preserving maps and the subsets of completely positive, decomposable, and superpositive qudit maps. Recently, it was shown that the volume of the positive but not completely positive, trace-preserving Pauli maps is twice as large as the volume of the Pauli channels \cite{Jagadish}. The geometry of Gaussian quantum channels with respect to the Bures-Fisher metric was considered by Monras and Illuminati \cite{Monras2010}.

In this paper, we derive the volumes of the positive and completely positive, trace-preserving Pauli maps. In order to achieve this, we calculate the Hilbert-Schmidt line and volume elements for the Choi-Jamio{\l}kowski states that correspond to the Pauli maps. In the next step, we consider the volumes of the entanglement breaking channels and the channels that can be generated using the time-local generator. Finally, we also find the relative volumes of the CP and P-divisible Pauli channels. Our results can be interpreted as a probability that a randomly selected Pauli channel is entanglement breaking, divisible, or has positive eigenvalues.

\section{Pauli maps and Pauli channels}

Consider the Pauli map, which is the most general bistochastic qubit map \cite{King,Landau} defined by
\begin{equation}\label{Pauli}
\Lambda[\rho]=\sum_{\alpha=0}^3p_\alpha\sigma_\alpha\rho\sigma_\alpha
\end{equation}
with the Pauli matrices $\sigma_0=\mathbb{I}_2$ and 
\begin{equation}
\sigma_1=\begin{pmatrix} 0 & 1 \\ 1 & 0 \end{pmatrix},\quad
\sigma_2=\begin{pmatrix} 0 & - i \\  i & 0 \end{pmatrix},\quad
\sigma_3=\begin{pmatrix} 1 & 0 \\ 0 & -1 \end{pmatrix}.
\end{equation}
Let us characterize these maps in terms of their eigenvalues $\lambda_\alpha$. The eigenvalue equation reads
\begin{equation}
\Lambda[\sigma_\alpha]=\lambda_\alpha\sigma_\alpha
\end{equation}
with $\lambda_0=1$. There is a simple relation between $p_\alpha$ and $\lambda_\alpha$; namely,
\begin{equation}
\lambda_\alpha=p_0+2p_\alpha-\sum_{\beta=1}^3p_\beta,\qquad\alpha=1,2,3.
\end{equation}
To map qubits into qubits, the Pauli map has to be positive, which is the case if and only if
\begin{equation}\label{P}
|\lambda_\alpha|\leq 1.
\end{equation}
Now, if the extended map $\oper_N\otimes\Lambda$ transforms quantum states into quantum states, where $\oper_N$ is the $N$-dimensional identity map, then $\Lambda$ is completely positive and is called the {\it Pauli channel}. Recall that a Pauli map is completely positive if and only if its eigenvalues satisfy the Fujiwara-Algoet conditions \cite{Fujiwara}
\begin{equation}\label{CP}
|1\pm\lambda_3|\geq|\lambda_1\pm\lambda_2|.
\end{equation}

\section{Geometry of the Pauli maps}

To analyze the geometry of the Pauli maps, we use the Choi-Jamio{\l}kowski isomorphism \cite{Choi,JAMIOLKOWSKI1972275} to express quantum channels as quantum states. Indeed, to the Pauli map $\Lambda$, there corresponds the unique Choi matrix
\begin{equation}
\begin{split}
\rho_\Lambda:&=\frac{1}{4}\sum_{i,j=0}^1|i\>\<j|\otimes\Lambda[|i\>\<j|]\\&
=\frac 14 \begin{pmatrix}
1+\lambda_3 & 0 & 0 & \lambda_1+\lambda_2 \\
0 & 1-\lambda_3 & \lambda_1-\lambda_2 & 0 \\
0 & \lambda_1-\lambda_2 & 1-\lambda_3 & 0 \\
\lambda_1+\lambda_2 & 0 & 0 & 1+\lambda_3
\end{pmatrix},
\end{split}
\end{equation}
whose eigenvalues are $p_\alpha$ from eq. (\ref{Pauli}). We equip the space of quantum maps with the metric $g=\frac 14 \mathrm{diag}(1,1,1)$ induced by the Hilbert-Schmidt line element
\begin{equation}\label{HS}
\der s^2:=\Tr(\der\rho_\Lambda^2)=\frac 14 (\der\lambda_1^2+\der\lambda_2^2+\der\lambda_3^2).
\end{equation}
The associated volume element has a simple form of
\begin{equation}\label{dV}
\der V:=\sqrt{g}\der\lambda_1\der\lambda_2\der\lambda_3=
\frac 18 \der\lambda_1\der\lambda_2\der\lambda_3
\end{equation}
If we integrate the volume element $\der V$ from eq. (\ref{dV}) over the regions $\mathcal{C}_{\rm PT}$ and $\mathcal{C}_{\rm CPT}$
given by ineq. (\ref{P}) and (\ref{CP}), respectively, we obtain the volumes
\begin{equation}
V(\mathcal{C}_{\rm PT})=1,\qquad V(\mathcal{C}_{\rm CPT})=\frac 13.
\end{equation}
of the positive, trace-preserving Pauli maps (PT) and the Pauli channels (CPT). Note that there are twice as many completely positive than positive but not completely positive maps. This reproduces the result from ref.  \cite{Jagadish}, where the authors proposed a measure to determine the volumes of PT and CPT Pauli maps.

\begin{figure}[ht!]
  \includegraphics[width=0.4\textwidth]{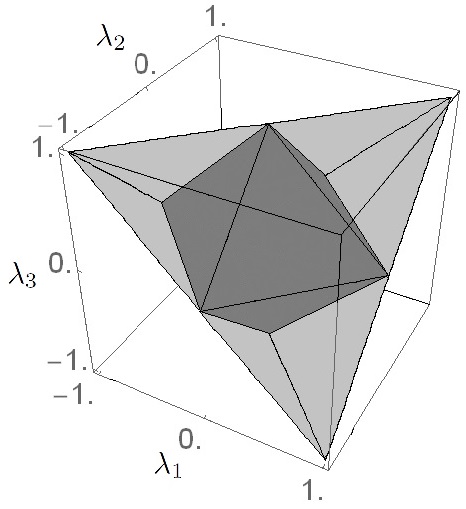}
\caption{
A graphical representation of the range of eigenvalues $\lambda_1,\ \lambda_2,\ \lambda_3$ that correspond to the Pauli channels (light grey) and the entanglement breaking Pauli channels (dark grey).}
\label{CPT_EBC}
\end{figure}

An important subclass of the Pauli channels are the entanglement breaking channels (EBC), whose extension $\oper_N\otimes\Lambda$ maps any state into a separable one. It was shown that $\Lambda$ is entanglement breaking if and only if \cite{qubitEBC}
\begin{equation}\label{EBC}
\sum_{\alpha=1}^3|\lambda_\alpha|\leq 1.
\end{equation}
Integrating the Pauli channels over the region $\mathcal{C}_{\rm EBC}$ defined by ineq. (\ref{EBC}) leads to the following volume,
\begin{equation}
V(\mathcal{C}_{\rm CPT}\cap\mathcal{C}_{\rm EBC})=\frac 16.
\end{equation}
In Fig. \ref{CPT_EBC}, we show the ranges of $\lambda_\alpha$ that correspond to the Pauli channels and the entanglement breaking Pauli channels, respectively. The complete positivity region is a tetrahedron (light grey), and the entanglement breaking region is an inscribed octahedron (dark grey). Observe that four of the octahedron's faces are coplanar with the tetrahedron's faces. The volume of the octahedron is half the volume of the tetrahedron.

Among their many applications, quantum channels are used to describe the dynamics of open quantum systems. Indeed, continuous time-evolution is provided with the use of time-parametrized families of quantum channels referred to as {\it dynamical maps}. As the simplest example of a Pauli dynamical map, one usually considers the Markovian semigroup $\Lambda(t)$, which is the solution of the master equation
\begin{equation}\label{MS}
\dot{\Lambda}(t)=\mathcal{L}\Lambda(t),\qquad\Lambda(0)=\oper,
\end{equation}
with the Gorini-Kossakowski-Sudarshan-Lindblad generator \cite{GKS,L}
\begin{equation}\label{L}
\mathcal{L}[\rho]=\frac 12 \sum_{\alpha=1}^3\gamma_\alpha(\sigma_\alpha\rho \sigma_\alpha-\rho),
\end{equation}
where $\gamma_\alpha\geq 0$.
To include non-Markovian effects, one introduces the time-local generator $\mathcal{L}(t)$, which has the same form as $\mathcal{L}$ in eq. (\ref{L}) but with time-dependent $\gamma_\alpha(t)$ that are not necessarily positive. Now, $\mathcal{L}(t)$ generates the dynamical map $\Lambda(t)$ with the eigenvalues	
\begin{equation}
\lambda_\alpha(t)=\exp[\Gamma_0(t)-\Gamma_\alpha(t)],
\end{equation}
where $\Gamma_\alpha(t)=\int_0^t\gamma_\alpha(\tau)\der\tau$ and $\gamma_0(t)=\sum_{\alpha=1}^3\gamma_\alpha(t)$. Another way to generalize Markovian semigroup master equation (\ref{MS}) is through the memory kernel master equation
\begin{equation}\label{K}
\dot{\Lambda}(t)=\int_0^tK(t,\tau)\Lambda(\tau)\der\tau,\qquad\Lambda(0)=\oper,
\end{equation}
with the non-local memory kernel $K(t,\tau)$. Note that the solutions of eq. (\ref{K}) can have negative eigenvalues $\lambda_\alpha(t)$.

\begin{figure}[ht!]
  \includegraphics[width=0.4\textwidth]{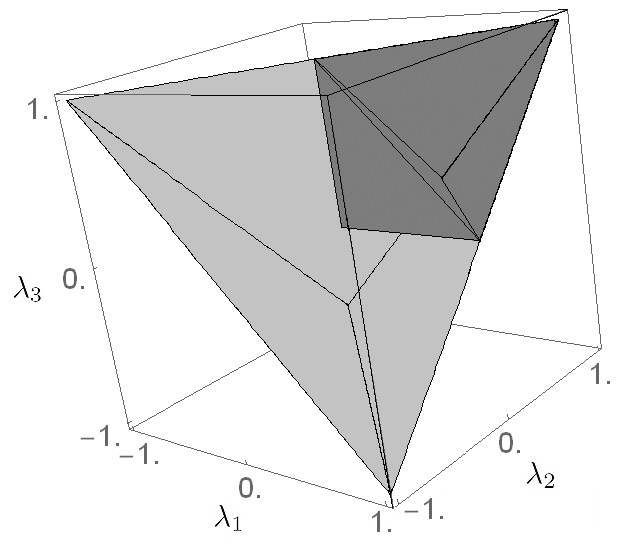}
\caption{
A graphical representation of the range of the eigenvalues $\lambda_1,\ \lambda_2,\ \lambda_3$ that describe the Pauli channels (light grey) and their subclass that is reachable with time-local generators (dark grey).}
\label{CPT_TLG}
\end{figure}

At this point, we make an important observation. Any Pauli channel $\Lambda$ can be obtained from a Pauli dynamical map $\Lambda^\prime(t=t_\ast)$ if and only if its eigenvalues $\lambda_\alpha$ are positive. The condition $\lambda_\alpha\geq 0$ corresponds to the region of integration $\mathcal{C}_{\rm TLG}$. Such channels are reachable with time-local generators (TLG), and their relative volume with respect to all Pauli channels is equal to
\begin{equation}
\frac{V(\mathcal{C}_{\rm CPT}\cap\mathcal{C}_{\rm TLG})}{V(\mathcal{C}_{\rm CPT})}=\frac{3}{16}.
\end{equation}
Meanwhile, the total volume of the positive maps reachable with time-local generators is
\begin{equation}
V(\mathcal{C}_{\rm PT}\cap\mathcal{C}_{\rm TLG})=\frac 18.
\end{equation}
Fig. \ref{CPT_TLG} shows the ranges of $\lambda_\alpha$ that correspond to the Pauli channels and the Pauli channels reachable with time-local generators. The complete positivity region is a tetrahedron (light grey), and the region of channels with positive eigenvalues is an inscribed triangular bipyramid (dark grey). Observe that two of the biparymid's faces are coplanar with the tetrahedron's faces. The volume of the triangular biparymid is $3/16$-th of the volume of the tetrahedron.
The remaining part,
\begin{equation}
1-\frac{V(\mathcal{C}_{\rm CPT}\cap\mathcal{C}_{\rm TLG})}{V(\mathcal{C}_{\rm CPT})}=\frac{13}{16},
\end{equation}
is the relative volume of the Pauli channels that are reachable only with the memory kernel $K(t,\tau)$.

\begin{figure}[ht!]
  \includegraphics[width=0.4\textwidth]{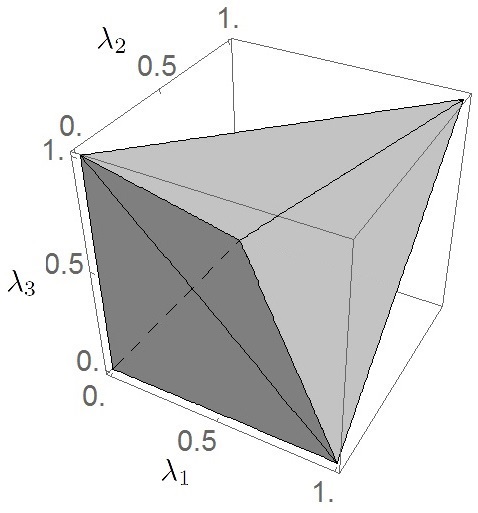}
\caption{
A graphical representation of the range of the eigenvalues $\lambda_1,\ \lambda_2,\ \lambda_3$ that describe the Pauli channels achievable with a time-local generator (light grey) and their entanglement breaking subclass (dark grey).}
\label{TLG_EBC}
\end{figure}

Additionally, one could ask what part of the Pauli channels reachable with time-local generators are entanglement breaking. It is straightforward to show that the corresponding volume ratio is
\begin{equation}
\frac{V(\mathcal{C}_{\rm CPT}\cap\mathcal{C}_{\rm TLG}\cap\mathcal{C}_{\rm EBC})}{V(\mathcal{C}_{\rm CPT}\cap\mathcal{C}_{\rm TLG})}=\frac 13.
\end{equation}
The associated ranges of $\lambda_\alpha$ are presented in Fig. \ref{TLG_EBC}. 
The triangular bipyramid (light grey) corresponds to the channels achievable with a time-local generator, and the inscribed tetrahedron (dark grey) corresponds to the entanglement breaking region. Three of the tetrahedron's faces lie on the same planes as the triangular bipyramid's faces. The volume of the tetrahedron is a third of the volume of the triangular biparymid. The figure that is a combination of Figs. 1--3 has recently been plotted in \cite{Davalos}, where divisibility of qubit channels is considered.

\begin{figure}[ht!]
  \includegraphics[width=0.45\textwidth]{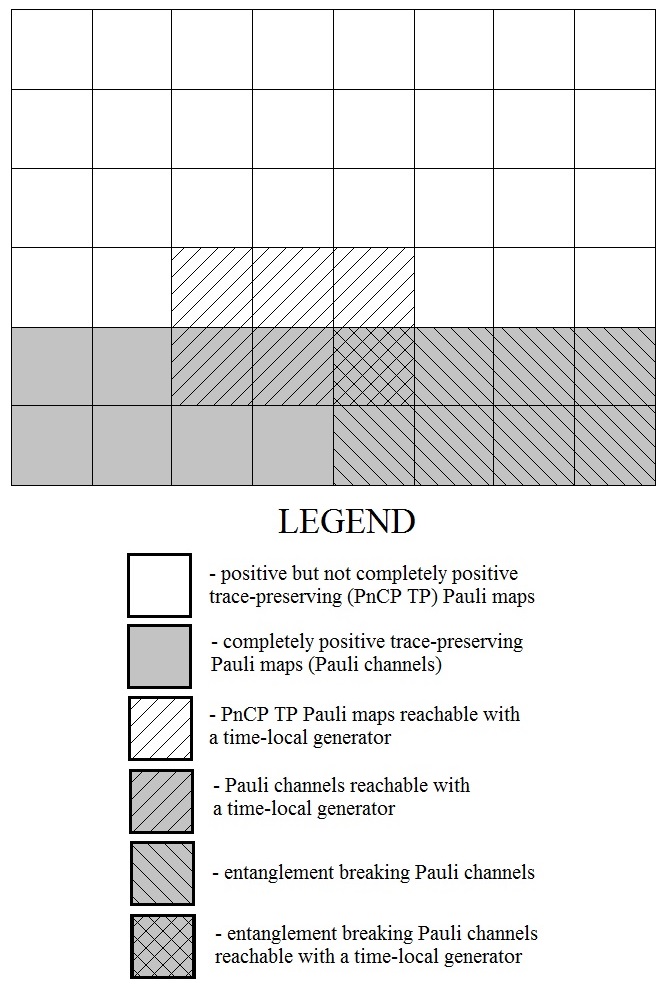}
\caption{A quantitative representation of the volumes of various Pauli maps. One square covers the region of $1/48$.}
\label{pie_chart}
\end{figure}

Let us present the main results of this section in one picture. In Fig. \ref{pie_chart}, the rectangle composed of $8\times 6$ identical squares represents the total volume of all positive, trace-preserving Pauli maps. The white region corresponds to the positive but not completely positive Pauli maps, whereas the grey region is associated with the Pauli channels (completely positive, trace-preserving Pauli maps). The single-hatched regions are the volumes of the entanglement breaking Pauli channels and the positive Pauli maps that are reachable with time-local generators, respectively (consult the legend in Fig. \ref{pie_chart} for details). Note that some of the regions overlap.

\section{P and CP-divisibility}

A special property of a quantum channel $\Lambda$ is its divisibility. Namely, $\Lambda$ is P-divisible if and only if it can be decomposed into
\begin{equation}
\Lambda=V\Lambda^\prime,
\end{equation}
where $\Lambda^\prime$ is a quantum channel and $V$ denotes a (non-unitary) positive, trace-preserving map. If $V$ is instead a quantum channel, then $\Lambda$ is said to be CP-divisible. In open quantum systems, divisibility of quantum dynamical maps is used to characterize Markovianity. Indeed, if $\Lambda(t)=V(t,s)\Lambda(s)$ for $s\leq t$ is CP-divisible, then it represents the Markovian evolution. A dynamical map that is only P-divisible is called {\it weakly non-Markovian} \cite{witness2,ChManiscalco}. Moreover, for invertible Pauli dynamical maps, weak non-Markovianity is equivalent to the lack of information backflow from the system to the environment \cite{Filip2}.

In the case of Pauli channels, divisibility is fully-characterized by the eigenvalues $\lambda_\alpha$. In particular, the Pauli channel is P-divisible if and only if \cite{Cirac}
\begin{equation}
\lambda_1\lambda_2\lambda_3\geq 0,
\end{equation}
which defines the P-divisibility region $\mathcal{C}_{\rm P-div}$.
The necessary and sufficient condition for CP-divisibility of the invertible Pauli channel reads \cite{Wolf}
\begin{equation}
0<\lambda_1\lambda_2\lambda_3\leq(\min_\alpha\lambda_\alpha)^2.
\end{equation}
We denote the associated region by $\mathcal{C}_{\rm CP-div}$.
Note that non-invertible Pauli channels do not contribute to the volume due to $\der\lambda_\alpha=0$.
The relative volumes of the P and CP-divisible Pauli channels with respect to all Pauli channels amount to
\begin{align}
\frac{V(\mathcal{C}_{\rm CPT}\cap\mathcal{C}_{\rm P-div})}{V(\mathcal{C}_{\rm CPT})}&=\frac{3}{4},\\
\frac{V(\mathcal{C}_{\rm CPT}\cap\mathcal{C}_{\rm CP-div})}{V(\mathcal{C}_{\rm CPT})}&=\frac 38,
\end{align}
respectively. Interestingly, $V(\mathcal{C}_{\rm CPT}\cap\mathcal{C}_{\rm P-div})/V(\mathcal{C}_{\rm CPT})$ has the same value as the relative volume of the Pauli channels simulable with a one-qubit mixed state environment \cite{Arvind}. Analogical calculations can be carried over for the Pauli channels that are reachable with time-local generators. The results are
\begin{align}
\frac{V(\mathcal{C}_{\rm CPT}\cap\mathcal{C}_{\rm TLG}\cap\mathcal{C}_{\rm P-div})}{V(\mathcal{C}_{\rm CPT}\cap\mathcal{C}_{\rm TLG})}&=1,\\
\frac{V(\mathcal{C}_{\rm CPT}\cap\mathcal{C}_{\rm TLG}\cap\mathcal{C}_{\rm CP-div})}{V(\mathcal{C}_{\rm CPT}\cap\mathcal{C}_{\rm TLG})}&=\frac{1}{2},\label{CPD}
\end{align}
where eq. (\ref{CPD}) agrees with Ref. \cite{Filippov}. 
Observe that half of the P-divisible maps are CP-divisible, even if we restrict our attention to the Pauli channels with positive eigenvalues. The geometrical regions representing the P and CP-divisible Pauli channels have already been plotted in \cite{Davalos}, and they are very involved.

\begin{figure}[ht!]
  \includegraphics[width=0.45\textwidth]{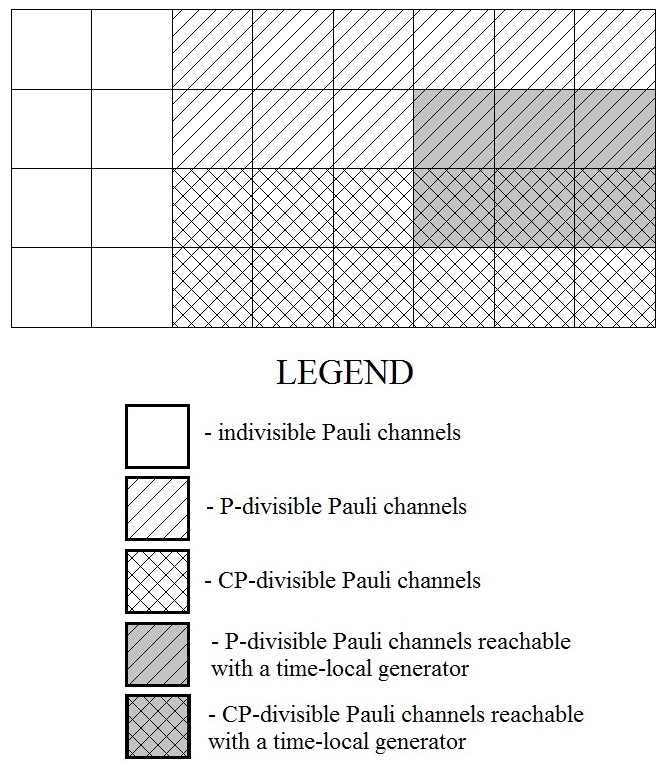}
\caption{A quantitative representation of the volumes of divisible Pauli channels. One square covers $1/32$ of the volume of the Pauli channels.}
\label{pie_chart2}
\end{figure}

Fig. \ref{pie_chart2} summarizes the results of this section. The rectangle composed of $8\times 4$ identical squares represents the volume of all Pauli channels. The unhatched region corresponds to the indivisible Pauli channels, whereas the grey region is associated with the Pauli channels that are reachable with time-local generators. The single and double-hatched regions are the relative volumes of the P and CP-divisible Pauli channels, respectively. More details can be found in the legend in Fig. \ref{pie_chart2}. Note that some regions overlap.

\section{Conclusions}

In this paper, we analytically calculate the volumes of the positive, trace-preserving Pauli maps and Pauli channels. We use the Choi-Jamio{\l}kowski isomorphism and the Hilbert-Schmidt line element to introduce the metric in the space of Pauli maps. The associated volume element depends only on the map's eigenvalues. Next, we find the volumes of some special classes of the Pauli channels: entanglement breaking, reachable with time-local generators (having positive eigenvalues), as well as P and CP-divisible.

Our results allow one to check the probability that a randomly generated Pauli channel or positive map has desired properties. Also, they make us realize what a small percentage of Pauli channels is covered by the time-local generators and how important it is to further develop the memory kernel approach. Similarly, the ratio of CP and P-divisible channels with respect to indivisible Pauli channels can shed some light on the amount of dynamical maps describing Markovian and non-Markovian quantum evolutions.

There are several topics that require further study. It would be interesting to consider the volumes of more general quantum maps, like non-unital qubit maps or unital qudit maps. However, one is sure to encounter problems with determining the regions of integration. For example, the range of admissible eigenvalues for non-unital qubit channels or positive qudit channels are not known.

Recall that the geometry of a given space depends on the metric. An alternative choice to the metric induced by the Hilbert-Schmidt line element in eq. (\ref{HS}) is the one following from the Fisher-Rao line element \cite{Fisher}
\begin{equation}
\der s^2_{FR}:=\Tr(\der\sqrt{\rho_\Lambda})^2.
\end{equation}
Observe that the corresponding volume element
\begin{equation}
\der V_{FR}=\frac{1}{8\sqrt{p_0p_1p_2p_3}}
\der\lambda_1\der\lambda_2\der\lambda_3
\end{equation}
does not coincide with $\der V$ from eq. (\ref{dV}). Moreover, one can encounter technical issues with analytical integration of $\der V_{FR}$, so numerical analysis might be necessary.

\section*{Acknowledgements}

This paper was supported by the Polish National Science Centre project No. 2018/31/N/ST2/00250.

\bibliography{C:/Users/cynda/OneDrive/Fizyka/bibliography}

\begin{thebibliography}{10}
\providecommand{\url}[1]{\texttt{#1}}
\providecommand{\urlprefix}{URL }
\providecommand{\eprint}[2][]{\url{#2}}

\bibitem{Zyczkowski}
I.~Bengtsson and K.~\.{Z}yczkowski, \textit{Geometry of Quantum States: An
  Introduction to Quantum Entanglement},  Cambridge University Press, Cambridge
  2007.

\bibitem{Vitale}
V.~I. Ma'nko, G.~Marmo, F.~Ventriglia, and P.~Vitale, J. Phys. A: Math. Theor.
  \textbf{50}, 335302 (2017).

\bibitem{Sommers2}
H.-J. Sommers and K.~\.{Z}yczkowski, J. Phys. A: Math. Gen. \textbf{36},
  10115--10130 (2003).

\bibitem{Sommers}
H.-J. Sommers and K.~\.{Z}yczkowski, J. Phys. A: Math. Gen. \textbf{36},
  10083--10100 (2003).

\bibitem{Link2015}
V.~Link and W.~T. Strunz, J. Phys. A: Math. Theor. \textbf{48}, 275301 (2015).

\bibitem{Arvind}
G.~Narang and Arvind, Phys. Rev. A \textbf{75}, 032305 (2007).

\bibitem{Jung}
E.~Jung, M.-R. Hwang, Y.~H. Ju, D.~K. Park, H.~Kim, M.-S. Kim, and J.-W. Son,
  J. Phys. A: Math. and Theor. \textbf{41}, 045306 (2008).

\bibitem{Choi}
M.-D. Choi, Linear Algebra Appl. \textbf{10}, 285--290 (1975).

\bibitem{JAMIOLKOWSKI1972275}
A.~Jamio{\l}kowski, Rep. Math. Phys. \textbf{3}, 275--278 (1972).

\bibitem{Lovas}
A.~Lovas and A.~Andai, Rev. Math. Phys. \textbf{30}, 1850019 (2018).

\bibitem{Szarek2}
S.~J. Szarek, E.~Werner, and K.~\.{Z}yczkowski, J. Math. Phys. \textbf{49},
  032113 (2008).

\bibitem{Jagadish}
V.~Jagadish, R.~Srikanth, and F.~Petruccione, Phys. Rev. A \textbf{99}, 022321
  (2019).

\bibitem{Monras2010}
A.~Monras and F.~Illuminati, Phys. Rev. A \textbf{81}, 062326 (2010).

\bibitem{King}
C.~King and M.~B. Ruskai, IEEE Trans. Info. Theory \textbf{47}, 192--209
  (2001).

\bibitem{Landau}
L.~J. Landau and R.~F. Streater, Linear Algebra Appl. \textbf{193}, 107--127
  (1993).

\bibitem{Fujiwara}
A.~Fujiwara and P.~Algoet, Phys. Rev. A \textbf{59}, 3290 (1999).

\bibitem{qubitEBC}
M.~B. Ruskai, Rev. Math. Phys. \textbf{15}, 643--662 (2003).

\bibitem{GKS}
V.~Gorini, A.~Kossakowski, and E.~Sudarshan, J. Math. Phys. \textbf{17}, 821
  (1976).

\bibitem{L}
G.~Lindblad, Comm. Math. Phys. \textbf{48}, 119 (1976).

\bibitem{Davalos}
D.~Davalos, M.~Ziman, and C.~Pineda, Quantum \textbf{2}, 144 (2019).

\bibitem{witness2}
D.~Chru\'{s}ci\'{n}ski, C.~Macchiavello, and S.~Maniscalco, Phys. Rev. Lett.
  \textbf{118}, 080404 (2017).

\bibitem{ChManiscalco}
D.~Chru\'{s}ci\'{n}ski and S.~Maniscalco, Phys. Rev. Lett. \textbf{112}, 120404
  (2014).

\bibitem{Filip2}
D.~Chru\'{s}ci\'{n}ski and F.~A. Wudarski, Phys. Rev. A \textbf{91}, 012104
  (2015).

\bibitem{Cirac}
M.~M. Wolf and J.~I. Cirac, Commun. Math. Phys. \textbf{279}, 147 (2008).

\bibitem{Wolf}
M.~M. Wolf, J.~Eisert, T.~S. Cubitt, and J.~I. Cirac, Phys. Rev. Lett.
  \textbf{101}, 150402 (2008).

\bibitem{Filippov}
S.~N. Filippov, J.~Piilo, S.~Maniscalco, and M.~Ziman, Phys. Rev. A
  \textbf{96}, 032111 (2017).

\bibitem{Fisher}
R.~A. Fisher, Proc. Camb. Phil. Soc. \textbf{22}, 700 (1925).

\end{thebibliography}
\bibliographystyle{beztytulow2}

\end{document}